# Internal field induced enhancement and effect of resonance in Raman scattering of InAs nanowires

Jaya Kumar Panda, Anushree Roy
Department of Physics, Indian Institute of Technology Khragpur. Pin 721302. India

Achintya Singha
Department of Physics, Bose Institute, 93/1, Acharya Prafulla Chandra Road, Kolkata 700 009, India

Mauro Gemmi
Center for Nanotechnology Innovation @ NEST, Istituto Italiano di Tecnologia, Piazza S. Silvestro 12, I-56127 Pisa, Italy

Daniele Ercolani, Vittorio Pellegrini, Lucia Sorba
NEST-Istituto Nanoscienze-CNR and Scuola Normale Superiore, Piazza S. Silvestro 12, I-56127Pisa, Italy

**Abstract**

An internal field induced resonant intensity enhancement of Raman scattering of phonon excitations in InAs nanowires is reported. The experimental observation is in good agreement with the simulated results for the scattering of light under varying incident wavelengths, originating from the enhanced internal electric field in an infinite dielectric cylinder. Our analysis demonstrates the combined effect of the first higher lying direct band gap energy ($E_1$) and the refractive index of the InAs nanowires in the internal field induced resonant Raman scattering. Furthermore, the difference in the relative contribution of electro-optic effect and deformation potential in Raman scattering of nanowires and bulk InAs over a range of excitation energies is discussed by comparing the intensity ratio of their LO and TO phonon modes.

# I. Introduction

Nanoparticles couple uniquely with electromagnetic (EM) radiation. Local field enhancement of visible EM radiation by nanometer-sized metal particles is well studied in the literature.[1,2,3] There are quite a few reports on the strong coupling of the local field of metal particles (under EM radiation) with the dielectric system in metal-dielectric conjugated nanostructures and also with dielectric nanostructures in metallic matrix.[4,5,6] Though, not very commonly found in the literature, the optical resonance in dielectric nanostructures, alone, by EM radiation has also been discussed in the literature.[7,8] The incident monochromatic electromagnetic radiation creates a distribution of electric fields within the dielectric nanostructures. The resonance occurs for a certain combination of the size of particles, wavelength of incident radiation and relative refractive index of the medium. The resultant effect is manifested in large enhancements in Raman scattering intensity from the dielectric particles, when the excitation wavelength is close to its EM eigenmode.

EM energy inside an irradiated cylinder was first calculated analytically by Ruppin.[9] A strong enhancement of local/internal field in dielectric nanostructures, like silicon nanowires and nanocones of varying diameters was experimentally observed.[10] In a recent report,[11] resonant Raman scattering measurements were carried out to probe the electronic band structure near the $E_1$ gap of InAs NWs, which are mostly of wurtzite (WZ) phase. The variation of the integral intensity of the phonon modes with laser excitation energy reveals redshifted $E_1$ gap of the WZ NWs compared to bulk InAs, of zinc-blende (ZB) phase. However, the expected enhancement in the intensity of the phonon mode due to large internal EM field inside the irradiated NWs, was not discussed. Probing strong coupling of radiation with semiconductor nanostructures is technologically relevant, since the development of ultrahigh-Q and ultralow-threshold lasers often exploits this unique characteristic of semiconductors.[12]

We also remark that the study on phonon dynamics reveals other interesting phenomena associated with NWs. The dipole-allowed Raman scattering by TO phonons in a polar semiconductor is a result of deformation-potential (DP) induced electron-phonon interaction, whereas the Raman scattering by LO phonon involves both DP-induced and electro-optic (EO) interactions. The latter is the Fröhlich-type interaction, coupling interband electronic energy states by the electric field of the LO phonon.[13] Thus, the comparison of the intensity ratio of these two phonon modes has the potential to shed light on the above effects in these semiconductors.

In this paper we report internal field enhanced spontaneous Raman scattering from InAs nanowires (NWs) and its dependence on the wavelength of excitation laser source. The enhancement is modeled by simulating the scattering intensity of infinitely long cylinder of InAs of a fixed diameter under different wavelengths of excitation. Furthermore, the difference in the intensity ratio of LO and TO phonon modes in InAs bulk and NWs near the resonance excitation energy is addressed in view of the difference in their electronic structure. In section II of this article, we have briefly discussed the technique of sample growth, which we have followed. Other experimental details are also available in section II. Results on HRTEM and Raman scattering measurements on NWs are discussed in section III. Section IV(a) deals with variation of enhanced Raman scattering from InAs NWs for varying excitation wavelengths as well as a brief discussion explaining the observation. In section IV(b) we compare the intensity ratio of the LO and TO phonon modes for NWs and bulk InAs by resonance Raman spectroscopy. Section V summarizes our results with a few concluding remarks.

## II. Experimental Details

Aligned InAs NWs were grown on InAs substrate (111)B using chemical beam epitaxy technique at 425±5$^{o}$C, with metallorganic line pressures of 0.3 and 1.0 Torr for TMIn and TBAs, respectively.[14] The structure of the NWs was studied using Zeiss Ultraplus field emission gun scanning electron microscope (SEM) and Zeiss LIBRA 120 transmission electron microscope (TEM). For TEM and Raman measurements, samples were prepared by mechanically removing NWs from the substrate with a razor blade and then transferred onto 300 mesh copper TEM grids coated with 50 nm carbon films. Samples were placed on a Si substrate for Raman measurements. Raman measurements were carried out using a micro-Raman setup equipped with $Ar^{+}$-$Kr^{+}$ laser (Model 2018-RM, make Newport, USA) as the excitation light source, a spectrometer (model T64000, make JY, France) and a CCD detector. For resonance Raman measurements, spectra of both bulk and InAs NWs were recorded at ten different wavelengths of excitation between 457 nm and 647 nm.

## III. TEM and Raman scattering

In Fig. 1(a) and (b) we show a characteristic SEM image and a dark field TEM image of the NW, respectively. More details of the structural analysis of this sample has been discussed elsewhere.[15] The average diameter and length of the NWs are estimated after measuring 15 different NWs to be 44±2 nm and 2.0±0.2 μm, respectively. In order to measure the length fraction of WZ and ZB in the NW, we collected dark field TEM images of the NWs oriented in [2-1-10]. In this orientation the pattern of the WZ and ZB structures are distinguishable and the diffraction pattern of WZ and ZB do not superimpose. Therefore a dark field image shows different contrast for the two crystal structures. Then, the fraction of WZ to the total length of the NW is estimated in the sample to be 0.95.

Raman spectra of InAs bulk and NWs, recorded at 78 K using 514 nm as excitation wavelength, are shown in Fig. 2. The experimental data points are shown by + symbols. By deconvoluting the spectra (dashed lines in Fig. 2) with Lorentzian functions we have estimated the allowed phonon wavenumbers. For bulk InAs [111] , TO and LO phonons appear at 217.2 $cm^{-1}$ and 239.0 $cm^{-1}$ (as obtained by deconvoluting the spectrum by two Lorentzian functions for these two modes). For NWs in addition to the LO mode at 239 $cm^{-1}$ and the $E_1^H$ mode at 217.2 $cm^{-1}$, a new $E_2^H$ mode, a folded phonon mode of WZ phase, appear at 212 $cm^{-1}$.[11] Here we would like to mention that for NWs, the net fitted curve does not match very well over the spectral window between 225 and 235 $cm^{-1}$. It should be stressed that similar mismatch between the data points and net fitted spectrum is also observed for bulk InAs (Fig. 2) over the same spectral range. Thus, we believe that the possible spectral feature between TO and LO modes is not related to any unique characteristic of a NW (say, surface phonon modes). Such asymmetry in the LO mode at lower wavenumber can also be observed in Ref. [11] in the Raman spectrum of InAs NWs.

## IV(a). Internal field enhanced Raman scattering

To obtain the expression for the internal field in the NW under EM radiation, let us consider the NW to be an infinite cylinder of radius $R$ and dielectric constant $\varepsilon$, surrounded by a medium of dielectric constant $\varepsilon_m$. Given an EM radiation of wavelength λ and of a particular polarization, one has to solve the Helmohltz equation for a scalar function in cylindrical polar co-ordinates to obtain the scattered and internal fields. The scalar function ($\psi_n$), which satisfies the Helmohltz equation, is a generating function for vector harmonics of order *n*; $\bar{M}_n = \bar{\nabla}\times(\hat{e}_z\psi_n)$ and $\bar{N}_n = \frac{\bar{\nabla}\times\bar{M}_n}{k}$.[16] The fields can then be expressed in terms of the above cylindrical vector functions[16]

$$\begin{aligned} \bar{M}_n &= \bar{\nabla}\times[\hat{e}_z Z_n(kr)\exp(in\theta)], \\ \bar{N}_n &= \frac{1}{k}\bar{\nabla}\times\bar{M}_n. \end{aligned} \tag{1}$$

$r, \theta$, and z are cylindrical coordinates and $\hat{\varepsilon}_z$ is the unit vector in the z-direction. $Z_n(kr)$ represents a cylindrical Bessel or Hankel function. $J_n(k_1 r)$ with $k_1=2\pi m(\lambda)/\lambda$ is used for the estimation of the field inside the cylinder. $J_n(k_0 r)$ and $H_n(k_0 r)$ with $k_0=2\pi/\lambda$ are used for the incident and scattered waves outside the cylinder. $m(\lambda)$ is the real part of the refractive index of the wire. The internal field can be calculated for two independent polarization states of incident and scattered waves (refer to Fig. 3). The plane of incidence contains the incident direction and the axis of the wire. In general convention, TM mode is the one for which the electric field vector is in the plane of incidence, and for the TE mode the magnetic field vector is in the plane of incidence.[17] For these two polarization configurations, after expanding the expression for the electric field in cylindrical polar coordinates along with the proper continuity conditions, the internal electric field of the NW per unit length can be obtained. Hence, one yields the expression for the volume-averaged intensity of the modes per unit length as[10]

$$I_{\text{int}}^{TM} = E_0^2 \sum_{n=-\infty}^{n=\infty} |d_n|^2 [J_n^2(k_1R)] - J_{n-1}(k_1R)J_{n+1}(k_1R)] \tag{2}$$

$$I_{\text{int}}^{TE} = \frac{E_0^2}{2} \sum_{n=-\infty}^{n=\infty} |c_n|^2 \{[J_{n-1}^2(k_1R)] + J_{n+1}^2(k_1R) - J_n(k_1R)[J_{n-2}(k_1R)] + J_{n+2}(k_1R)]\} \tag{3}$$

The coefficients $c_n$ and $d_n$ are also functions of $\lambda$ and $R$. $E_0$ is the amplitude of the incident electric field. The enhanced electric field results in increase in inelastic scattering (Raman) intensity of the incident radiation. With the approximation that the scattered frequency is not very different from that of incident frequency, the ratio of the scattered intensity per unit volume ($I$) of the NW to that of bulk is given by

$$S(\lambda) = \left[\frac{I_{\text{NW}} / \text{V}_{\text{NW}}}{I_{\text{bulk}} / \text{V}_{\text{bulk}}}\right]_{\text{expt}} = \left[\frac{I^{TE,TM}(\lambda)}{I_\alpha(\lambda)}\right]^2 \tag{4}$$

where $\text{V}_{\text{NW}}$ and $\text{V}_{\text{bulk}}$ are the volumes of the NW and bulk probed by the radiation and $I_\alpha$ is the average intensity of radiation within the dispersive medium. The above expression for $S$ strongly depends on $k_1$ and $k_0$, and hence on λ and $m(\lambda)$. To simulate $S(\lambda)$, the values of real part of the refractive index for different values of $\lambda$ for bulk InAs are taken from Ref. [18]. The data points for $m$ over three different regions of wavelengths have been fitted with 2$^{\text{nd}}$ order polynomial equations. Using these known empirical relations for $m(\lambda)$,the variation of $S$ with incident wavelength ($\lambda$) for TM and TE modes for different radii of the wires are shown in Fig. 4. Strong enhancement (note the log scale in Fig. 4) in the intensity of the scattered signal is predicted for both modes, for certain combinations of the diameter of the NW and excitation wavelength. For larger diameters of the NW, a steeper resonance condition is achieved.

Raman spectra of bulk InAs, recorded for different wavelengths of excitations in ($(\bar{z}(x,x)z)$ scattering configurations are shown in Fig. 5(a). The same for NW of InAs recorded in TM configuration for varying wavelengths of excitations are shown in Fig. 5(b). It is to be noted that in the spectra of NWs, recorded at room temperature, the $E_2^H$ and $E_1^H$ modes could not be resolved. Therefore we have fitted each spectrum of NW over the spectral range between 200 and 225 $cm^{-1}$ by two Lorentzian functions for above two modes. For bulk InAs a single function for $E_1^H$ mode is used. In Fig. 6 we have plotted (symbols) the ratio of the intensity (*S*) of the $E_1^H$ TO mode for NWs to that of bulk InAs, as obtained experimentally. The solid line is the simulated *R* using Eqn. 5 for R= 22 nm with a shift by 10 nm (from 504 nm to 514 nm ≡48 meV) towards the higher wavelength to match the experimental data points. In order to rationalize this shift of 48 meV we recall that the solid lines correspond to the results of the simulations for the InAs nanostructure assuming the same electronic properties of bulk InAs, which is in a ZB phase. There are quite a few reports in the literature, where the electronic band structure of the WZ phase of InAs has been calculated.[19,20,21,22] In a recent article,[11] it was shown that the band gap related resonant Raman intensity of the TO mode of the InAs NW down-shifts by 110 meV from that of the corresponding bulk material due to the decrease in $E_1$ gap of the WZ phase of the former. In the inset to Fig. 6, the intensity of the TO mode of NW and bulk InAs, as measured by us, has been plotted using blue and green symbols for different laser excitation wavelengths. As expected from the electronic band structure and results reported in Ref.[11], we also observe the resonance of the TO mode of the NW and bulk InAs at 513 nm and 491 nm (difference ≡108 meV). Thus, if we take into account this 108 meV downshift of the resonance energy, we can conclude that the 48 meV downshift of the maxima in internal field induced enhancement of Raman scattering in NWs, as observed in Fig. 6, is not a manifestation of band-gap resonance alone. A crucial role is played by the material property, *m(λ),* in determining internal field induced Raman scattering for the dielectric NW.

## IV(b). Contributions of DP and EO effect in Raman scattering of NW

In Fig. 7 we compare the intensity ratio (ρ) of LO and TO modes for bulk and NW of InAs with varying incident excitation energy. It should be recalled that this ratio corresponds to relative contribution of deformation potential and electro-optic effect in the inelastic scattering (Raman) of light in polar semiconductor. For the excitation energy below the energy of the $E_1$ gap the relative contribution of DP and EO effect on intensity of the phonon modes is not expected to alter appreciably. Thus, the value of ρ is nearly same till 2.5 eV and 2.4 eV for bulk and NWs, as observed in Fig. 7. However, near resonance, the ratio increases with the increase of the excitation energy due to enhanced coupling of electric field of electromagnetic radiation with phonons. For the excitation energy well above the band gap, the ratio decreases for bulk InAs (shown in the inset of Fig. 7), while for NWs we find that there is a monotonous increase in ρ beyond the band gap, with further increase in excitation energy (till 2.8 eV). To explain this anomalous behavior of ρ in NWs, we refer to the electronic band structure of WZ and ZB phases of the NW as obtained from ab-initio calculation.[11] At the $E_1$ gap ($\Gamma \rightarrow A$) of the WZ phase, there are closely spread electronic energy states (within the range of 0.4 eV) in the valence band. This increases the range of excitation energy (2.4 eV to 2.8 eV in Fig. 7), over which the strong coupling of the electric field of electromagnetic radiation with the phonon mode can be obtained. However, at the $E_1$ gap ($\Gamma \rightarrow L$) of the ZB phase, such degeneracy in energy states is lifted. Hence a relatively sharp resonance in ρ is observed for the bulk system.

## V. Conclusions

In conclusion, we have demonstrated a strong resonance effect determined by the increase in electric field intensity of electromagnetic radiation in InAs NWs yielding a large enhancement in inelastic scattering of light from InAs NWs. We have also provided a classical explanation for the enhancement effect in our cylindrical NWs. These findings demonstrate how an efficient coupling of visible radiation with such semiconductor nanostructures can be achieved. The giant field enhancement in InAs NWs harness the potential of using these semiconductor nanostructures in future semiconductor nanotechnology, in sensor engineering and photonic devices, where efficient coupling with electromagnetic radiation is desirable. In addition, we have compared the relative contribution of DP and EO effect in NWs and bulk InAs for varying excitation energies.

**Acknowledgements**

AR and JKP thank Central Research Facility at IIT Kharagpur for the availability of Raman spectrometer.

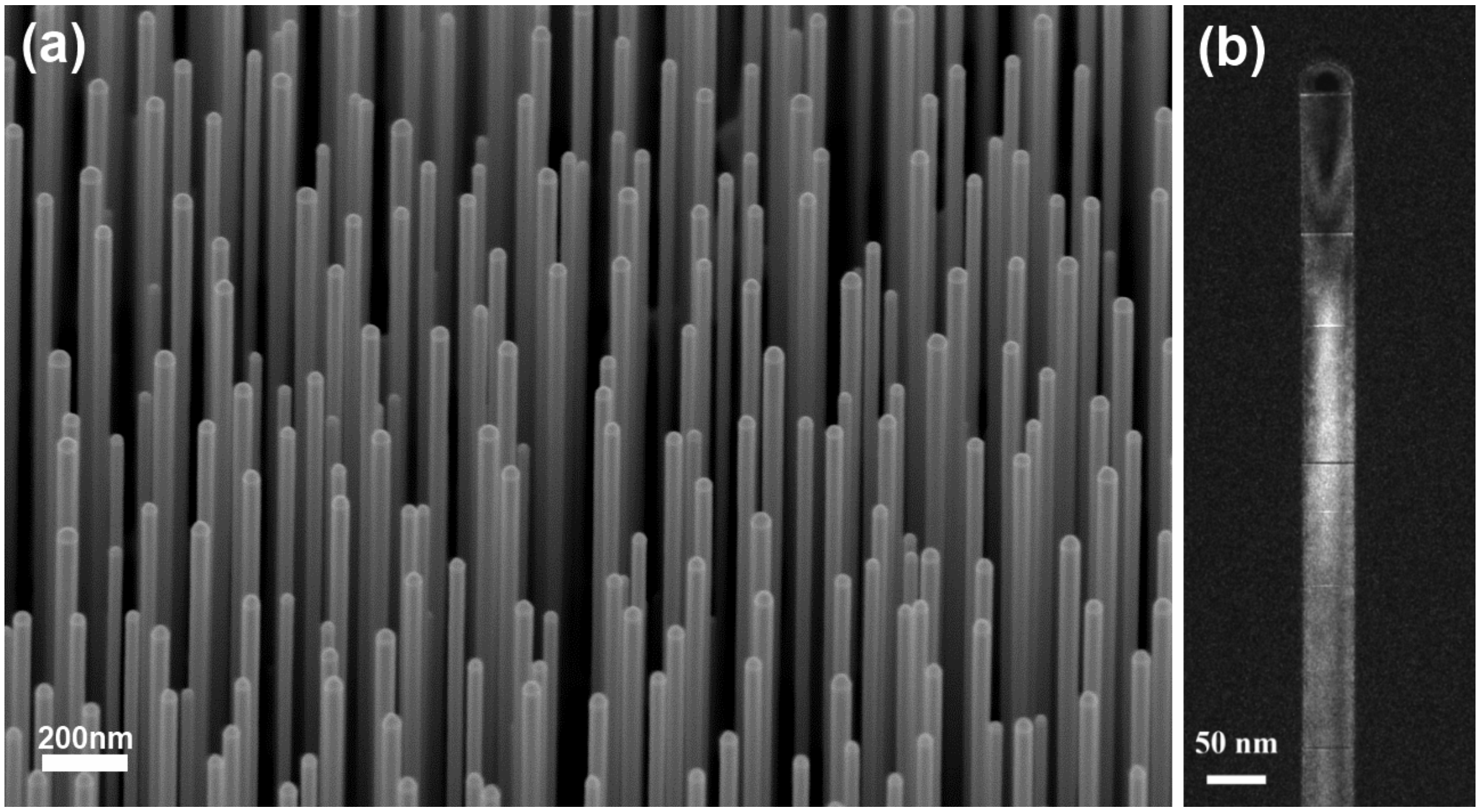

Fig. 1. (a) SEM micrograph of the as-grown sample; the substrate is tilted by an angle of 45°, b) TEM image of a typical single InAs NW.

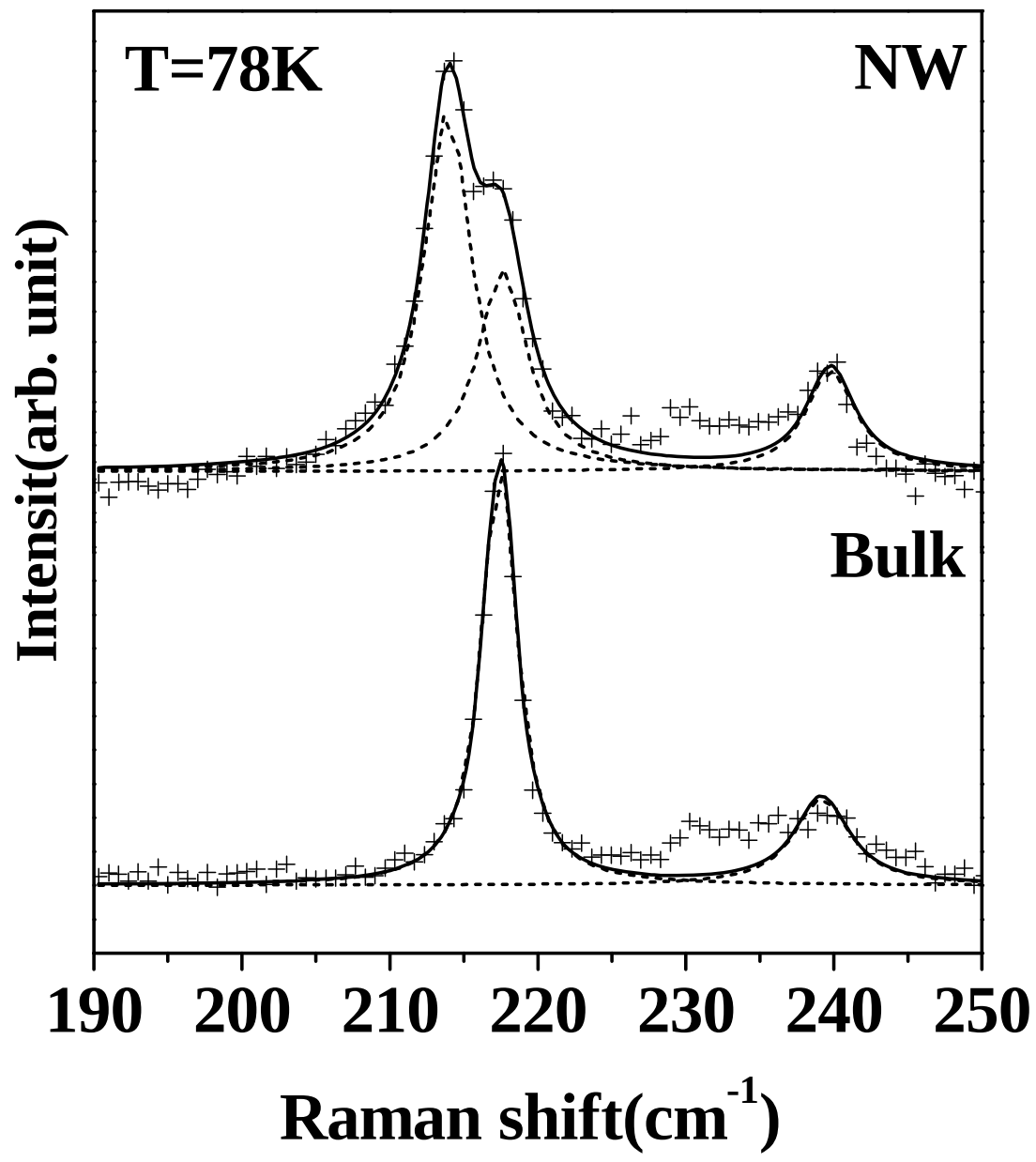

Fig. 2. Raman spectrum of bulk and NWs of InAs recorded at 78 K. Dotted lines are the deconvoluted spectra for TO and LO modes. The net fitted spectrum is shown by the solid line.

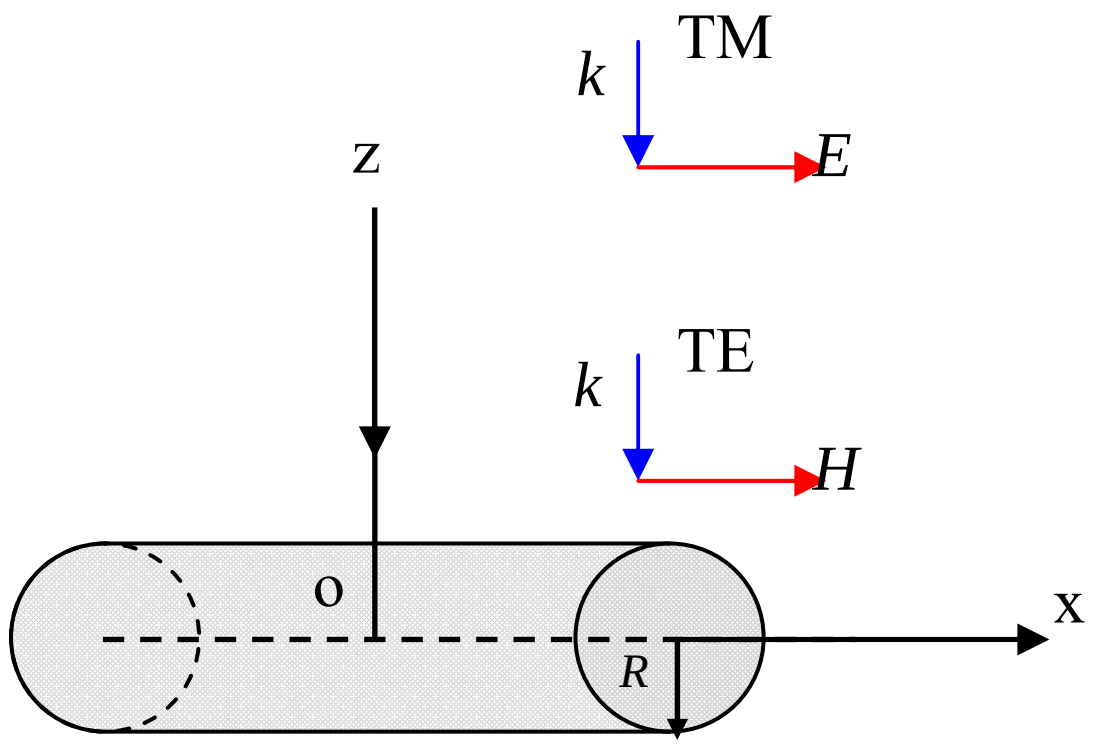

Fig. 3. Schematic diagram to show the possible polarization configurations of incident and scattered lights. x is the axis of the wire. xoz is the plane of incidence. The incident wavevector *k* (blue arrow), and the electric field or magnetic field (red arrows) vectors of the scattered light, corresponding to TM or TE mode, respectively, are shown by blue arrows.

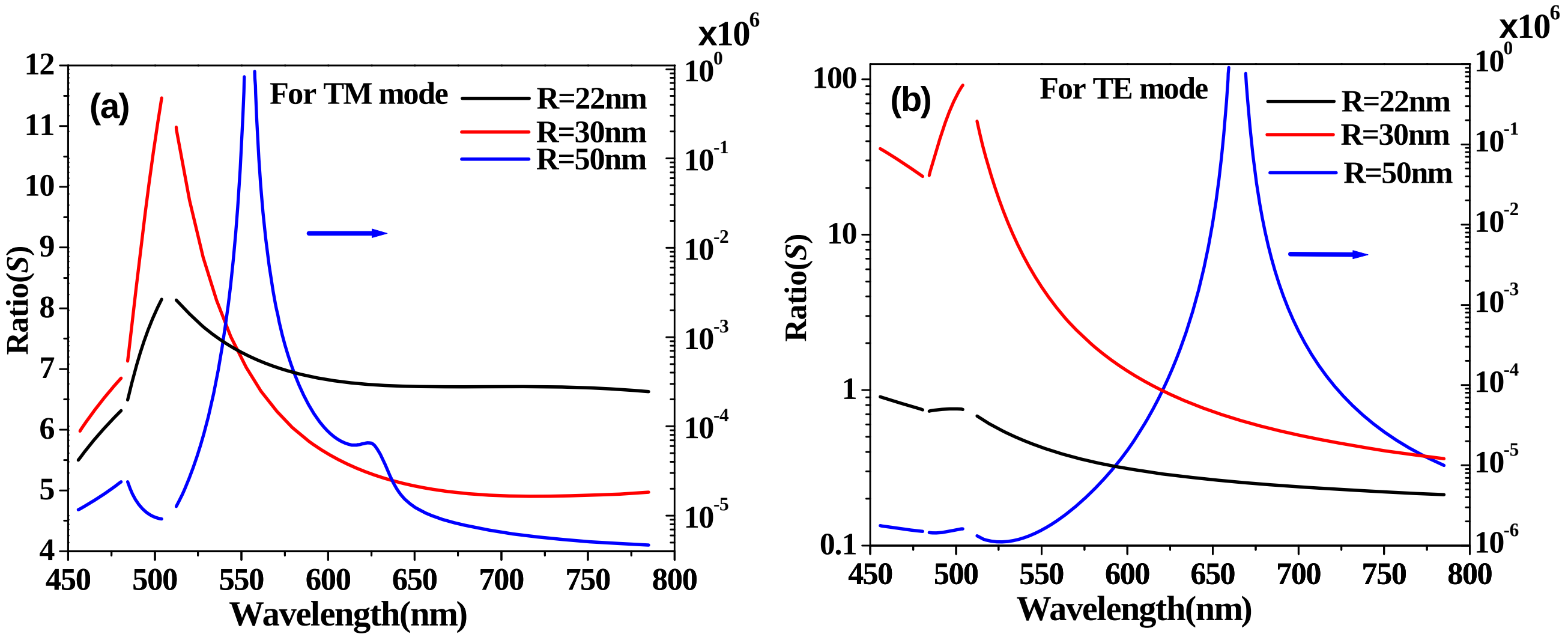

Fig. 4: Variation in the square of the ratio of scattered intensity per unit volume of the NW to that of bulk ($S(\lambda)$) with wavelength of radiation for different radii of InAs NWs for (a) TM mode and (b) TE mode of radiation.

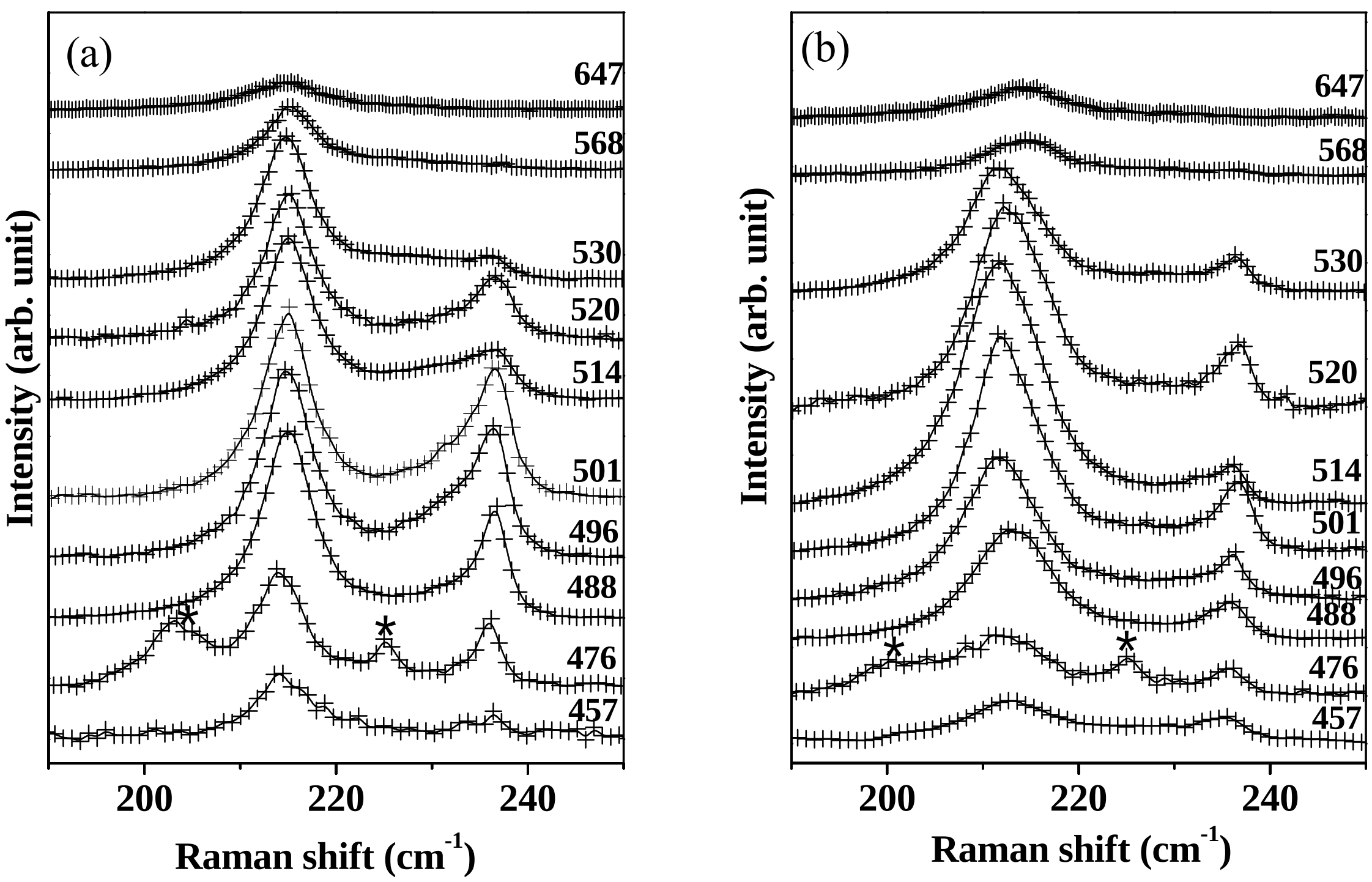

Fig. 5. Raman spectra of (a) bulk and (b) NWs of InAs, recorded at varying excitation wavelengths (in nm at right). The * marks are two unidentified lines observed in both bulk and NWs for 476 nm excitation wavelength.

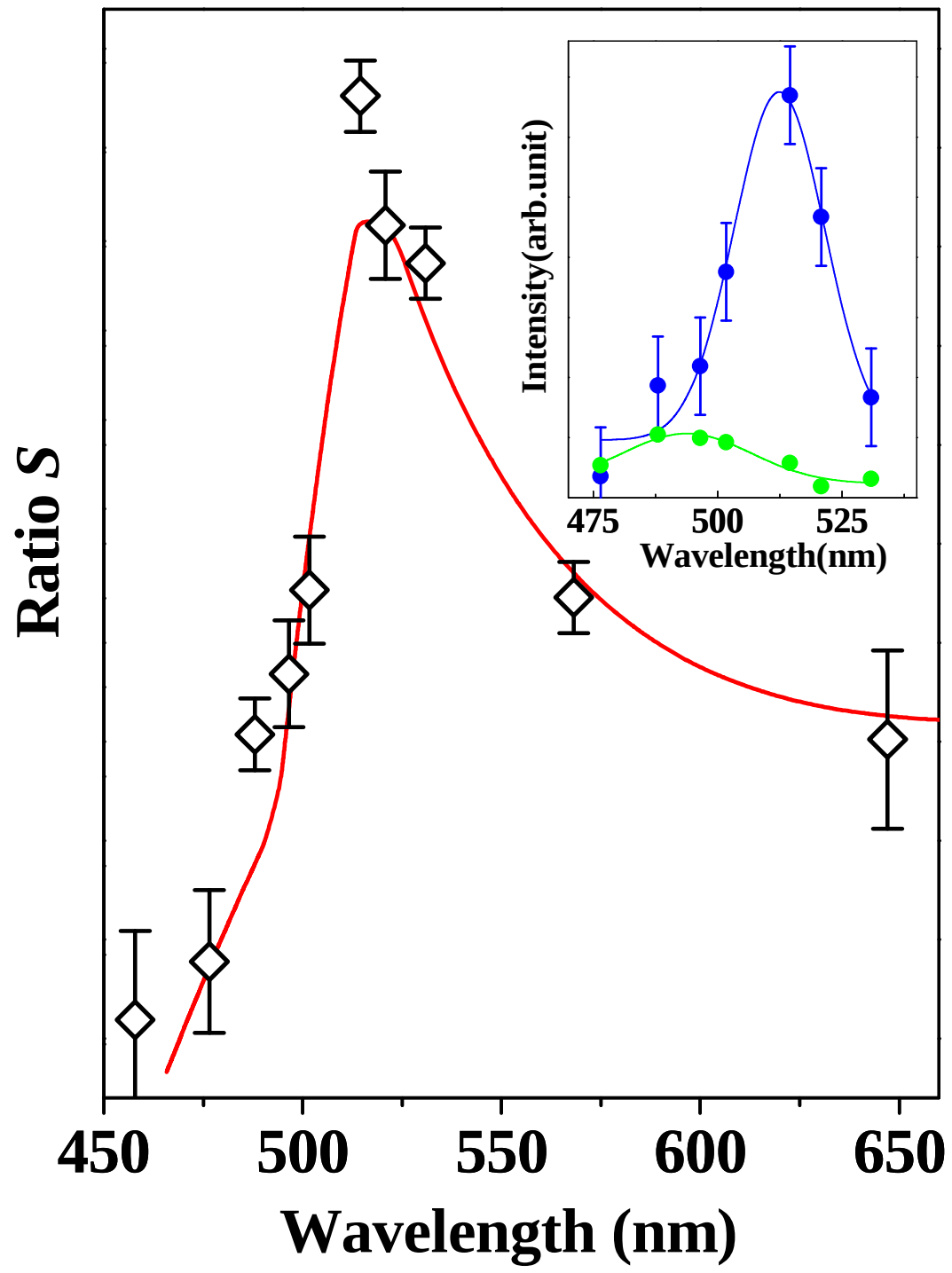

Fig. 6. Variation of *S* with wavelength. Experimental data points are shown by symbols. The solid lines are simulated curve for *S(λ)* as obtained from Eqn. 4. Inset of the figure shows the variation of integral intensity of $E_1^H$ TO modes of the InAs NWs (blue dots) and bulk (green dots) with wavelength of excitation.

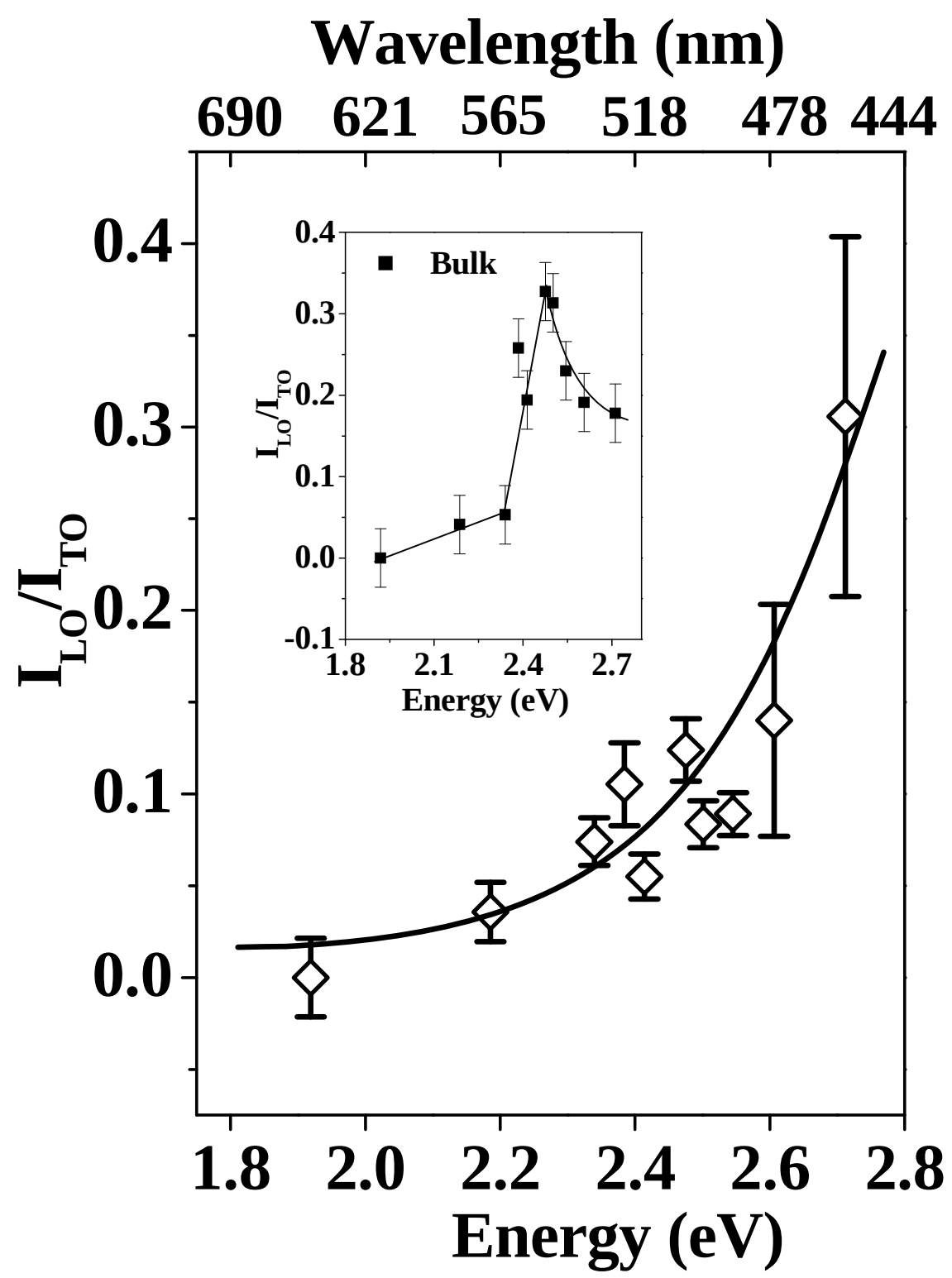

Fig. 7: Variation of intensity ratio of LO and TO modes with incident excitation energy in NWs. Inset of the figure shows the same for bulk InAs. The solid lines are the guide to the eye. Error bars at last two points are large because of low intensity of the TO mode.

[20] Z. Zanolli, F. Fuchs, J. Furthmuller, U. V. Barth and F. Bechstedt, Phys. Rev. B **75**, 245121 (2007).

[21] A. De and C. E. Pryor, Phys. Rev. B **81**, 155210 (2010).

[22] C. Thelander, P. Caroff, S. Plissard, A.W. Dey and K.A. Dick, Nano Lett. **11**, 2424 **(**2011)**.**